\documentclass[reprint,superscriptaddress,amsmath,amssymb,aps,prd,twocolumn,floatfix,nofootinbib]{revtex4-2}
\usepackage{amsmath, amssymb}
\usepackage{listings}
\usepackage{graphicx}
\usepackage{dcolumn}
\usepackage{bm,hyperref}
\usepackage{float,color}
\usepackage{xcolor}
\usepackage[normalem]{ulem}
\usepackage{tabularx}
\usepackage{mathtools} 
\usepackage{multirow}
\usepackage{amssymb}
\usepackage[caption=false]{subfig}
\usepackage{scrextend}
\usepackage{hyperref}

\begin{document}

\title{
Identifying intermediate mass binary black hole mergers in AGN disks using LISA
}
\author{Poulami Dutta Roy}
\affiliation{Chennai Mathematical Institute, Siruseri, 603103, India}\email{poulami@cmi.ac.in}
\author{Parthapratim Mahapatra}
\affiliation{School of Physics and Astronomy, Cardiff University, Queens Buildings, Cardiff, CF24 3AA, United Kingdom}
\author{Anuradha Samajdar}
\affiliation{Institute for Gravitational and Subatomic Physics (GRASP), Utrecht University, Princetonplein 1, 3584 CC Utrecht, The Netherlands}
\affiliation{Nikhef – National Institute for Subatomic Physics, Science Park 105, 1098 XG Amsterdam, The Netherlands}
\author{K. G. Arun}
\affiliation{Chennai Mathematical Institute, Siruseri, 603103, India}
\begin{abstract}

We show that Laser Interferometer Space Antenna can uniquely identify the sites of intermediate-mass binary black hole (IMBBH) mergers if they occur in Active Galactic Nuclei (AGN) disks with a gas density $\rho\geq10^{-12} \, {\rm g/cc}$ via measurement of dynamical friction effect in the gravitational waveform. We find that even a single observation of a gravitational wave source with a total mass of $10^3 M_{\odot}$ and a mass ratio of 2 at a luminosity distance of 3 Gpc is sufficient to confidently associate the merger to be in an AGN disk with a density $\sim 10^{-12} \, {\rm g/cc}$, as it allows estimation of the density with an error bar ${\cal O}(100\%)$.
This provides a new way of inferring AGN disk densities that complement traditional X-ray observations. Further, we find that neglecting the presence of environmental effects in the waveform models used for parameter estimation can bias the chirp mass, mass ratio and arrival time of a merger. If not corrected, this can significantly impact our ability to carry out multiband data analysis of IMBBHs that combines information from LISA and ground-based gravitational wave detectors.
\end{abstract}
\maketitle
\section{Introduction}
Active Galactic Nuclei disks (AGN disks) are excellent sites for rapid and efficient growth of intermediate mass black holes (IMBHs)~\cite{Madau:2001sc,Miller:2003sc,McKernan:2012rf,McKernan:2014oxa, Bartos:2016dgn, Stone:2016wzz,Mckernan:2017ssq}. Compared to other astrophysical environments, such as nuclear star clusters (NSC), which have densities ranging from $\sim 10^{-18} \rm g/cm^3$ to $\sim 10^{-15} \rm g/cm^3$~\cite{Chattopadhyay:2023pil}, and globular clusters (GC) with densities between $10^{-21}-10^{-18} \, {\rm g/cm^3}$~\cite{Antonini:2022vib}, the cores of AGN disks can have densities between $10^{-12}-10^{-6} \, {\rm g/cm^3}$ making them the densest astrophysical environments where such mergers occur~\cite{Fabj:2024kqs,Sirko:2003,Thompson:2005mf}. With its capability to detect IMBH binaries at high redshifts with high signal-to-noise ratios (SNRs), the planned Laser Interferometer Space Antenna (LISA) mission~\cite{LISA:2017pwj} could observe a population of intermediate-mass binary black hole (IMBBH) mergers occurring in AGN disks.

Measurement of orbital eccentricity and spin-induced precession are considered to be the standard diagnostics to establish the association of binary black hole (BBH) mergers in dense environments (see, for instance, Refs.~\cite{Mapelli:2020vfa,Mandel:2021smh} for reviews). However, these features cannot necessarily distinguish mergers in different types of dense environments, such as NSC and GC, from those in AGN disks, as the eccentricity (or spin) distribution of binaries from these environments is hard to resolve.  Arguably, one of the most robust indicators of a dense environment is the direct estimation of the density of the medium where the merger occurs.
As different environments that can host compact binary mergers have reasonably distinct average densities, such a measurement should aid in a better understanding of the merger sites. Therefore, it is pertinent to ask whether future gravitational wave (GW) observations could precisely measure the density of the nonvacuum environments in which mergers occur.

It is known that the phase evolution of gravitational wave signals of nonvacuum BBH mergers differs from their vacuum counterparts~\cite{Barausse:2014tra,Barausse:2014pra}.
The physical processes that can contribute to this dephasing include accretion onto the black hole (BH)~\cite{Caputo:2020irr}, dynamical friction due to mass segregation~\cite{Ostriker:1998fa,Kim:2007zb,Kocsis:2011dr,Macedo:2013qea}, etc. As this dephasing occurs at (effective) negative post-Newtonian (PN) orders (where low-frequency effects dominate in the early part of the inspiral), it is not measurable with the low-frequency sensitivities of the current or proposed future ground-based GW detectors~\cite{CanevaSantoro:2023aol, Roy:2024rhe}. However, we argue that mergers of IMBBHs with masses in the range $\sim10^2-10^4\,M_{\odot}$ with high SNRs, which will be observed by LISA up to very high redshifts ~\cite{Bellovary:2019nib, Miller:2008fi, Saini:2022hrs}, can give us a unique opportunity to precisely measure the density of the medium if the densities are above $\sim 10^{-12} \, {\rm g/cc}$. As this is the typical density of AGN disks~\cite{Thompson:2005mf, Fabj:2024kqs, Vajpeyi:2021qsw}, and few astrophysical environments can  reach such high densities, this implies LISA can confidently identify IMBBH mergers in AGN disks for a subset of the population at low redshifts ($z\leq 0.5$). 

\subsection{A quick review of past works}
References \cite{Barausse:2014tra,Barausse:2014pra} were among the first explorations of the effects of astrophysical environments on gravitational waveforms (see also Ref.~\cite{Barausse:2007dy}). These works studied a suite of physical effects that are related to the interaction of binary with the environment. They concluded that the impact of such environmental effects on inspiral and ringdown signals is generally small enough not to hinder precision inference, except possibly in the case of extreme mass-ratio inspirals~\cite{Barausse:2014tra}. Since then, several studies have investigated the detectability of these environmental effects across different frequency bands of the GW spectrum, covering the detectability of environmental effects in the context of supermassive ~\cite{Kocsis:2011dr,Meiron:2016ipr,DOrazio:2019fbq,Tiede:2023cje,Garg:2024oeu,Garg:2024qxq,Speri:2022upm,Zwick:2022dih,Derdzinski:2020wlw,Nouri:2023nss} intermediate-mass~\cite{Chandramouli:2021kts,Garg:2022nko,Boudon:2023vzl,Caputo:2020irr,Kadota:2023wlm,Rozner:2024vxo} and stellar-mass black holes~\cite{Cardoso:2019rou,Sberna:2022qbn,Cole:2022ucw,Zwick:2024yzh,Roy:2024rhe,CanevaSantoro:2023aol,Chen:2020lpq,Caputo:2020irr,Coogan:2021uqv}.

The recent LIGO-Virgo detection of the BBH merger GW190521~\cite{LIGOScientific:2020iuh,LIGOScientific:2020ufj}, whose binary constituents lie in the pair-instability mass gap~\cite{Belczynski:2016jno,Woosley:2021xba}, along with the report of an associated optical transient candidate~\cite{Graham:2020gwr}, has given a significant boost to those models that predict BBH mergers in AGN disks. 
AGN disks typically have high escape speeds, allowing them to retain a large fraction of the remnant BHs by preventing their escape due to gravitational recoil~\cite{Mahapatra:2021hme,Mahapatra:2022ngs,Mahapatra:2024qsy}.
Hence, AGN disks can easily grow BHs to the observed masses rapidly via accretion and repeated mergers~\cite{Tagawa:2019osr,Yang:2019cbr} and produce electromagnetic (EM) transients via interaction of the component or remnant BH with the surrounding gas~\cite{Bartos:2016dgn}. A population of BHs with component masses lying in the upper mass gap has also been reported in non-LVK GW transient catalogs~\cite{Mehta:2023zlk, Nitz:2021uxj}. This lends further support to this paradigm, though a confident association of this population with AGN disks remains challenging to establish with the current dataset.

The planned LISA observatory~\cite{LISA:2017pwj} will observe the early stages of GW190521-like binaries with low SNRs if they merge at a luminosity distance of 1 or 2 Gpc~\cite{Toubiana:2021iuw,Sberna:2022qbn}. Since environmental effects are dominant at low frequencies, LISA has the ability to detect them in systems such as GW190521~\cite{Toubiana:2021iuw}. Further, Ref.~\cite{Sberna:2022qbn} studied how well the motion of such a binary around the central supermassive black hole (SMBH) can be tracked via careful modelling of the frequency-domain GW signal and by examining various modulations in the signal. They argued that these measurements could, in turn, be mapped onto the parameters of the AGN disk.

More recently, Ref.~\cite{Copparoni:2025jhq} investigated the effect of stochastic gas torques on the waveforms of asymmetric binaries in the LISA band. By performing numerical simulations, they studied the systematics induced due to simple analytic models on LISA's ability to infer binary and environment parameters using intermediate- and extreme-mass-ratio inspirals in AGN disks. Using numerical modelling of compact binary evolution in AGN disks and incorporating the expected mass and spin distributions from AGN disk mergers, Ref.~\cite{Vajpeyi:2021qsw} argued that properties of AGN disks could be inferred when a sufficiently large number of such candidates is available. Finally, Ref.~\cite{CanevaSantoro:2023aol} recently used BBHs in the first GW transient catalog~\cite{GWTC-1} and set limits on the density of the medium in which they merged. As a follow-up, Ref.~\cite{Roy:2024rhe} studied the detectability of BBHs in nonvacuum environments using vacuum BBH templates in the context of ground-based detectors and investigated possible biases in tests of general relativity (GR) when these effects are not accounted for (see Ref.~\cite{Garg:2024qxq} for a study in the LISA context). They found that the detectability of nonvacuum mergers is not compromised even if vacuum templates are used. However, tests of GR could be biased if the density of the medium is extremely high ($\sim 10^{7}\,{\rm g/cc}$), though such a scenario is highly unlikely from an astrophysical perspective.

\subsection{Present work}
In this work, we explore the possibility of detecting imprints of environment in the GW signals emitted by IMBBH mergers in AGN disks~\cite{McKernan:2012rf, McKernan:2014oxa,Bartos:2016dgn}. Due to mass segregation~\cite{Morris:1993zz}, heavier BHs are drawn towards the central SMBH. Any merger in AGN disks will be impacted by dynamical friction caused by the gas-rich medium~\cite{Barausse:2014pra}. The extent to which the dynamics of a BBH are affected by the AGN disk will depend on the average density of gas in the disk. In this context, we ask whether the dephasing, caused by dynamical friction in the gravitational waveform from IMBBH mergers, is measurable, and if so, how accurately we can estimate the density of the ambient medium using LISA observations.
On the lower end of the IMBBH mass range ($\lesssim 10^3 M_{\odot}$), next-generation ground-based observatories such as Einstein Telescope (ET) and Cosmic Explorer (CE) will also observe these mergers. In those cases, combining data from both LISA and ET/CE for the same source, referred to as multibanding, can improve the estimation of background density. We quantify the improvement in the density estimation due to multibanding of IMBBH signals.

We find that IMBBH mergers in the LISA band, at a fiducial luminosity distance of 3 Gpc, can facilitate precise measurements of the ambient density if the density of AGN disks is $\geq {\cal O} (10^{-12}) \, {\rm g/cc}$. 
Further, we quantify the systematic bias on the estimation of time of coalescence, chirp mass, and symmetric mass ratio of IMBBH mergers in AGN disks when environmental effects are not included in the  gravitational waveform model during parameter estimation. 
We find that neglecting dynamical friction can lead to severe biases and affect our ability to perform multiband GW astronomy with LISA. Regarding systematic biases, our findings complement a similar conclusion for GW190521-like binaries reported in Ref.~\cite{Sberna:2022qbn}. Furthermore, our method of directly estimating the density of AGN disks complements the proposal in Ref.~\cite{Vajpeyi:2021qsw}, with the important difference that our inference requires only loud single events, rather than a population of events.

This paper is organized as follows: In Section II, we explain the structure of the leading-order contribution from dynamical friction in the phase of the gravitational waveform and its dependence on BBH parameters.
We also briefly explain the fundamentals of the Fisher matrix formalism used to obtain bounds on the parameters, including environmental density. In Section III, we discuss the main results of the paper: (i) the bound obtained by using LISA-only and multibanding of LISA and ET on the density of the surrounding environment for IMBBHs, and (ii) the systematic bias on the binary parameters when neglecting dynamical friction, followed by the conclusion in Section IV.

\section{Analysis set up}
\subsection{Gravitational waveforms with environmental effect}
In presence of astrophysical environments, the dynamics of a compact binary differs from that in vacuum. These environments could be distribution of masses or gas, which causes dynamical friction (DF), accretion of gas onto the  BH or the presence of a perturbing third body  that can  induce anomalous acceleration  of the center of mass. Among these effects, DF is expected to induce the most significant observable change in the GW phase for IMBBH mergers in AGN disks with high gas density. Therefore, we focus solely on the effect of DF on the GW phasing of IMBBHs\footnote{It is possible that the BHs involved in the merger could be accreting and/or that the binary  experiences acceleration of the CM due to the central SMBH. We have checked a few representative cases, and the inclusion of these effects does not change our main results. For example, the fractional error on density (see Sec.~\ref{sec:LISA_error}) of $10^{-10}$ $\rm g/cm^3$ remains of the order of $\sim 10^{-3}$ for $\rm M=1500 \,M_{\odot}, q=5$ binary when computed from only dynamical friction and including both dynamical friction and accretion into the GW phase.}. 
Our treatment of the phasing, which accounts for DF, closely follows those of Refs.~\cite{Chandrasekhar:1943ys,Ostriker:1998fa,Kim:2007zb,Toubiana:2020drf, CanevaSantoro:2023aol}.

We consider these binaries to be moving in quasi-circular orbits and account for the impact of DF only in the energy flux as an additional channel of energy loss, assuming that the orbital energy is the same as for vacuum inspirals. Using energy balance, the GW phasing evolution can then be computed under the adiabatic approximation. The leading-order contribution of the DF effect to the frequency-domain GW phasing is derived using the stationary phase approximation and is provided in Eq.~(4) of Ref.~\cite{Toubiana:2020drf}, which reads
\begin{align}
    \phi_{\rm DF} = - \rho \frac{25 \pi (3\eta -1) M_c^2}{739328\, \eta^2} \gamma_{\rm DF}\left[\pi M_c (1+z) f\right]^{-16/3}.
    \label{eq:DF}
\end{align}
In the above equation, $\eta= \frac{q}{(1+q)^2}$, where $q$ is the ratio of the primary mass to the secondary mass of the binary, called the symmetric mass ratio, and $M_c$ is the source-frame chirp mass, defined as $\rm{M_c} = \rm{M}  \eta^{3/5}$, with $\rm{M}$ being the binary’s total mass in the source-frame~\footnote{{ Due to cosmological expansion, the observed detector frame mass is $(1+z)$ times the source-frame mass. Throughout the paper, we quote the source-frame mass of the IMBBHs. Note that while computing the lower cut-off frequency through Eq.~(\ref{eq:f_low}), the reshifted chirp mass must be used, as that is what is  observed by the detector.}}, $z$ is the cosmological redshift of the source. The factor $\gamma_{\rm DF}$ in Eq.~\ref{eq:DF} is given by $\gamma_{\rm DF} = -247 \, \log \left(\tfrac{f}{f_{\rm DF}}\right) -39 + 304 \, \log (20) +38 \, \log \left(\tfrac{3125} {8}\right)$, where $f_{\rm DF}=\tfrac{c_s}{22 \pi M}$, and $c_s$ is the speed of sound in the astrophysical environment with density $\rho$. To get a representative value of speed of sound, following \cite{Fabj:2024kqs}, we select $10^{-10}$ g/$\rm cm^3$ as characteristic density of AGN disk which corresponds approximately to the speed of sound being 0.01 times the speed of light. Here, we have assumed $c_s=0.01 c$, where c is the speed of light for all cases.

Assuming a flat $\Lambda$-CDM model of our universe with  $\Omega_{M} = 0.3065$, $\Omega_\Lambda = 0.6935$ and $h = 0.6790$ with $\rm{H_0} =100\,h$ (km/s)/Mpc (\cite{Planck:2015fie}), the luminosity distance is related to the redshift by the relation
\begin{equation}
\rm{d_L} (z) = \frac{(1+z) }{H_0} \int_0^z \frac{dz'}{\sqrt{\Omega_M (1+z')^3 + \Omega_\Lambda}}.
\label{eq:redshift}
\end{equation}

We employ the {\tt IMRPhenomD}~\cite{Husa:2015iqa, Khan:2015jqa} waveform model as the vacuum GR waveform. {\tt IMRPhenomD} is a frequency-domain phenomenological model of the GW signal that captures the inspiral, merger, and ringdown phases of non-precessing (aligned-spin) black hole binaries in quasi-circular orbits. 
To incorporate the effect of DF in the gravitational waveform model, we modify the {\tt IMRPhenomD} waveform model by adding $\phi_{\rm DF}$ from Eq.~(\ref{eq:DF}) to its phase. The modified {\tt IMRPhenomD} waveform schematically reads as
\begin{align}
    \Tilde{h}_{\rm env}(f) = \mathcal{A}(f) \, e^{i (\Phi (f) + \phi_{\rm DF} (f))}\,, 
    \label{eq:env}
\end{align}
where $\mathcal{A}(f)$ and $\Phi(f)$ are the amplitude and phase of the {\tt IMRPhenomD} waveform model. We will use $\Tilde{h}_{\rm env}(f)$ as a nonvacuum waveform model for BBHs due to DF.

Neglecting precessional features in the waveform could have an impact on our estimates. However, intuitively, the effect of precession  is expected to be subdominant, as precession is a high-order PN effect  that will have negligible correlation with the density parameter. Nevertheless, a future study should confirm this and quantify the impact of precession on these estimates.

\subsection{Assessing the measurability of the density parameter}
Next, we want to assess the measurability of the density of the medium in which the IMBBHs are merging using LISA observations. A rigorous analysis of this problem would involve simulating a signal and injecting it into synthetic LISA data to obtain the posterior distribution  of the density parameter through a Bayesian analysis. Given the computational burden of this approach, here we use the Fisher information matrix to obtain error estimates on the density parameter using LISA. This will serve as a rough estimate of what can be expected and should be verified in the future using more robust parameter estimation techniques.

In the limit of large SNRs, and assuming the detector noise is stationary and Gaussian, the Fisher information matrix~\cite{Cramer46,Rao45,Cutler_Flanagan} (see \cite{Vallisneri07} for caveats) can be used to obtain the $1\sigma$ estimates of the statistical errors on the parameters due to the detector noise. Formally, the Fisher information matrix is the noise-weighted inner product of the derivatives of the frequency-domain waveform with respect to the binary parameters of interest, evaluated at the true values of these parameters. More precisely, the Fisher matrix is defined as
\begin{equation}
    \Gamma_{ab}=2\int_{f_{\rm low}}^{f_{\rm up}}\,\frac{{\tilde h}_{,a}{\tilde h}^{*}_{,b}+{\tilde h}_{,b}{\tilde h}^{*}_{,a}}{S_n(f)} df,
\end{equation}
where commas denote partial differentiation of the waveform $\Tilde{h}(f)$ with respect to various parameters $\theta^a$, the asterisk denotes complex conjugation, and $S_n(f)$ the noise power spectral density (PSD) of the detector of interest. The lower and upper limits of integration denote the low- and high- frequency cutoffs used in the analysis (see below for details). The SNR, which quantifies the strength of the signal in detector data, is defined using the Fourier transform of the signal $\Tilde{h}(f)$ as
\begin{eqnarray}
    {\rm SNR}^2 = 4 \int_{f_{\rm low}}^{f_{\rm up}} \frac{|\tilde{h}(f)|^2}{S_n (f)} df \,.
\end{eqnarray}
The $1\sigma$ error bar on a particular parameter is obtained by inverting the Fisher matrix and taking the square-root of the diagonal entries of the inverse matrix, known as the variance-covariance matrix $\Sigma_{ab}$. Specifically, the statistical error on a parameter $\theta^a$ is given by 
\begin{equation}
    \Delta \theta^a = \sqrt{\Sigma_{\rm{aa}}},  
\end{equation}
where $\Sigma_{\rm{ab}} = (\Gamma_{\rm{ab}} + \Gamma^0_{\rm{ab}})^{-1}$, $\Gamma^0_{\rm{ab}}$ is the prior matrix, and repeated indices are not summed over. We impose Gaussian priors on the phase of coalescence and the component spins, namely $\Gamma^0_{\phi_c \phi_c} = 1/\pi^2$ and $\Gamma^0_{\chi_1 \chi_1} = \Gamma^0_{\chi_2 \chi_2} = 1$, which helps to partially remedy the ill-conditionedness of the Fisher matrix, thereby making the numerical inversion of the matrix more reliable.

Our focus here is on the space-based future gravitational wave detector LISA, which will be sensitive in the mHz regime, where the effective $-$5.5PN order term due to dynamical friction will play a dominant role in the binary dynamics~\footnote{{Due to the presence of the logarithmic term in $\gamma$ in Eq.(~\ref{eq:DF}) this is not a $-$5.5PN term formally.}}. We do not account for the orbital motion of LISA, as it is not expected to play any significant role in the estimation of dynamical parameters such as the ones we are interested in. The lower cut-off frequency is chosen such that the GW signal from the inspiraling binary lasts for four years prior to its merger (or prior to leaving the LISA band), but is not lower than the low-frequency limit of the LISA noise PSD, which is $10^{-4}$ Hz.
Hence, the lower cut-off frequency is chosen as~\cite{BBW05a}
\begin{equation}
\begin{split}
   & f_{\rm low}  = {\rm max} \bigg[10^{-4},\\
   &\, 4.149 \times 10^{-5} \Big[ \frac{{\rm M_c (1+z)}}{10^6 M_{\odot}} \Big]^{-5/8}\left(\frac{\rm T_{obs}}{1\,{\rm yr}}\right)^{-3/8}\bigg],
    \label{eq:f_low}
\end{split}  
\end{equation}
where ${\rm T_{obs}=4 \,{\rm yrs}}$ is the observation time in LISA. 
The upper cut-off frequency of LISA is chosen to be 0.1 Hz.

The LISA noise PSD has contributions from both the instrumental noise and the galactic confusion noise, the latter being attributed to unresolved galactic white dwarf binaries. The instrumental noise PSD, given in \cite{Babak:2017tow}, is divided by a factor of 2 to account for summation over two independent frequency channels \cite{Robson:2018ifk}, while the background confusion noise in the low-frequency regime $f \lesssim 1$mHz is modeled through an analytical expression given in \cite{Mangiagli:2020rwz} for a four-year observation period of LISA. The upper cut-off frequency is the frequency at which the characteristic amplitude $2 \sqrt{f} |\tilde{h}(f)|$ of the GW signal is lower than that of the detector's noise amplitude spectral density by a maximum of 10\%. The intermediate-mass BBHs are considered at a prototypical luminosity distance ($\rm{d_L}$) of 3 Gpc for LISA. As errors vary inversely with SNR, which in turn varies roughly as $1/d_L$, the errors reported here can be approximately extrapolated to any luminosity distance.
\subsection{Multibanding}
 In addition to computing the bound on density obtained from an IMBBH signal observed in LISA, we also study the effect of {\em multibanding}. Multibanding refers to the observation of a signal (and subsequent parameter inference) by combining information  from two different bands of the GW spectrum. In the case of IMBBHs with total mass $\lesssim 10^3 M_{\odot}$, LISA will observe the early inspiral, but their mergers will fall within the frequency band of the ground-based detectors. Combining information from two different frequency bands will improve the parameter estimation. The implications of multibanding  have been discussed through multiple works in  the literature (see, for example, \cite{Nair:2015bga, Vitale:2016rfr, Sesana:2016ljz,Datta:2020vcj,Gupta:2020lxa}). 

Due to our interest in measuring a low-frequency effect, we consider the Einstein Telescope (ET) as a representative of future ground-based detector having a lower cutoff frequency of 1Hz. We have used the ET PSD in Ref.~\cite{Hild:2010id}, along with the upper cutoff frequency corresponding to the frequency at which the characteristic amplitude $2 \sqrt{f} |\tilde{h}(f)|$ of the GW signal is lower than that of the detector noise amplitude spectral density by at most 10\% . 
Using the Cosmic Explorer noise PSD~\cite{CE:2019iox} would not significantly alter our conclusions with regard to multibanding.

To obtain the bound on density by performing a Fisher analysis in the context of multibanding, we add the two Fisher matrices obtained from LISA and ET and then invert the  resulting total matrix. Hence, the total covariance matrix is given by
\begin{equation}
    \Sigma^{\rm combined} = (\Gamma^{\rm LISA} + \Gamma^{\rm ET} + \Gamma^0)^{-1} .
\end{equation}

In our analysis, we consider the following 8-dimensional parameter space:
\begin{equation}
    \theta = \{\rm{ln}\, \rm{d_L},t_c,\phi_c,\rm{ln}\, M_c, \rm{ln}\,\eta, \chi_1,\chi_2, \Bar{\rho}\},
\end{equation}
where $\chi_{1,2}$ denote the dimensionless spin parameters of the binary components, and $\rm{d_L}$ is the luminosity distance of the binary. The term $t_c$, called the time of coalescence, is a kinematical quantity closely related to the time of arrival of the signal in the detector, and $\phi_c$ denotes a constant phase. We find it convenient to deal with the dimensionless density parameter $\Bar{\rho} = \rho/\rho_0$, where $\rho_0$ is set to $10^{-10} \rm{g/cm^3}$  as a characteristic density. The choice of $\rho_0$, following Ref.~\cite{Toubiana:2020drf}, does not affect the final results and is motivated by the density  values used in our study, which range from $10^{-8}\, \rm{g/cm^3}$ to $10^{-12} \, \rm{g/cm^3}$ \cite{Fabj:2024kqs}.

\subsection{Systematic error computation}
In addition to computing the $1\sigma$ statistical error on $\Bar{\rho}$, we also  assess the potential systematic biases in parameter estimation that may arise if environmental effects, such as dynamical friction, are neglected. The systematic error in the estimation of a parameter is defined as the difference between its ``true" value, $\theta_a^{\rm T}$, and  its ``best fit" value, $\hat{\theta}_a$,  which corresponds to the peak of the recovered Gaussian probability distribution
\begin{equation}
    \Delta \theta_a^{\rm sys} = \theta_a^{\rm T} - \hat{\theta}_a.
\end{equation}

This can be computed using the Cutler-Vallisneri formalism \cite{Cutler:2007mi}. The {\it true} waveform model that describes the physical system, which in this study refers to a model that includes environmental effects in the waveform. An {\it approximate} waveform model, $\Tilde{h}_{\rm AP}$, in our case the vacuum waveform, is represented by the approximate amplitude $\mathcal{A}_{\rm AP}$ and approximate phase $\phi_{\rm AP}$
\begin{equation}
     \Tilde{h}_{\rm AP} = \mathcal{A}_{\rm AP} e^{i \phi_{\rm AP}} \,, 
     \end{equation}
and the true waveform $\Tilde{h}_{\rm T}$ differs from $\Tilde{h}_{\rm AP}$ in amplitude and phase by $\Delta \mathcal{A}$ and $\Delta\phi$, respectively, 
\begin{eqnarray}
    \Tilde{h}_{\rm T} = [\mathcal{A}_{\rm AP}+ \Delta \mathcal{A}] e^{i [\phi_{\rm AP} + \Delta \phi]} \,.
\end{eqnarray}
The systematic error can then be approximated as~\cite{Cutler:2007mi, Favata:2021vhw,Saini:2022igm}
\begin{equation}
    \Delta \theta^a_{\rm sys} \approx \Sigma^{ab} \Big([\Delta \mathcal{A} + i\mathcal{A}_{\rm AP} \Delta \phi] e^{i \phi_{\rm AP}} \Big\vert \partial_b \Tilde{h}_{\rm AP} \Big) \,,
\end{equation}
where $\Sigma_{ab}$ is the covariance matrix that is calculated using the approximate waveform, and $\Delta\phi = \phi_{\rm DF}$.  Repeated indices are summed over. We do not account for any corrections to the amplitude, i.e., $\Delta \mathcal{A} =0$. When computing the systematic bias, we consider the 7-dimensional parameter space, $\theta = \{\rm{ln}\, \rm{d_L},t_c,\phi_c,\rm{ln}\, M_c, \rm{ln}\,\eta, \chi_1,\chi_2\}$, which excludes the density parameter.

Given the high SNRs of the events (see below in Sec. IIIA) considered here, we expect our estimates will be representative of the results that will come from a fully Bayesian treatment of the problem. For instance, our Fisher matrix code results are in good agreement with the Bayesian estimates of \cite{Toubiana:2020drf}. However, a dedicated Bayesian study of the problem would help in putting our estimates on a firmer footing. We plan to undertake this as a follow up project.

\section{Results}
\begin{figure}[t]
    \centering
    \subfloat[]{\label{fig:bound}%
  \includegraphics[width=0.41\textwidth]{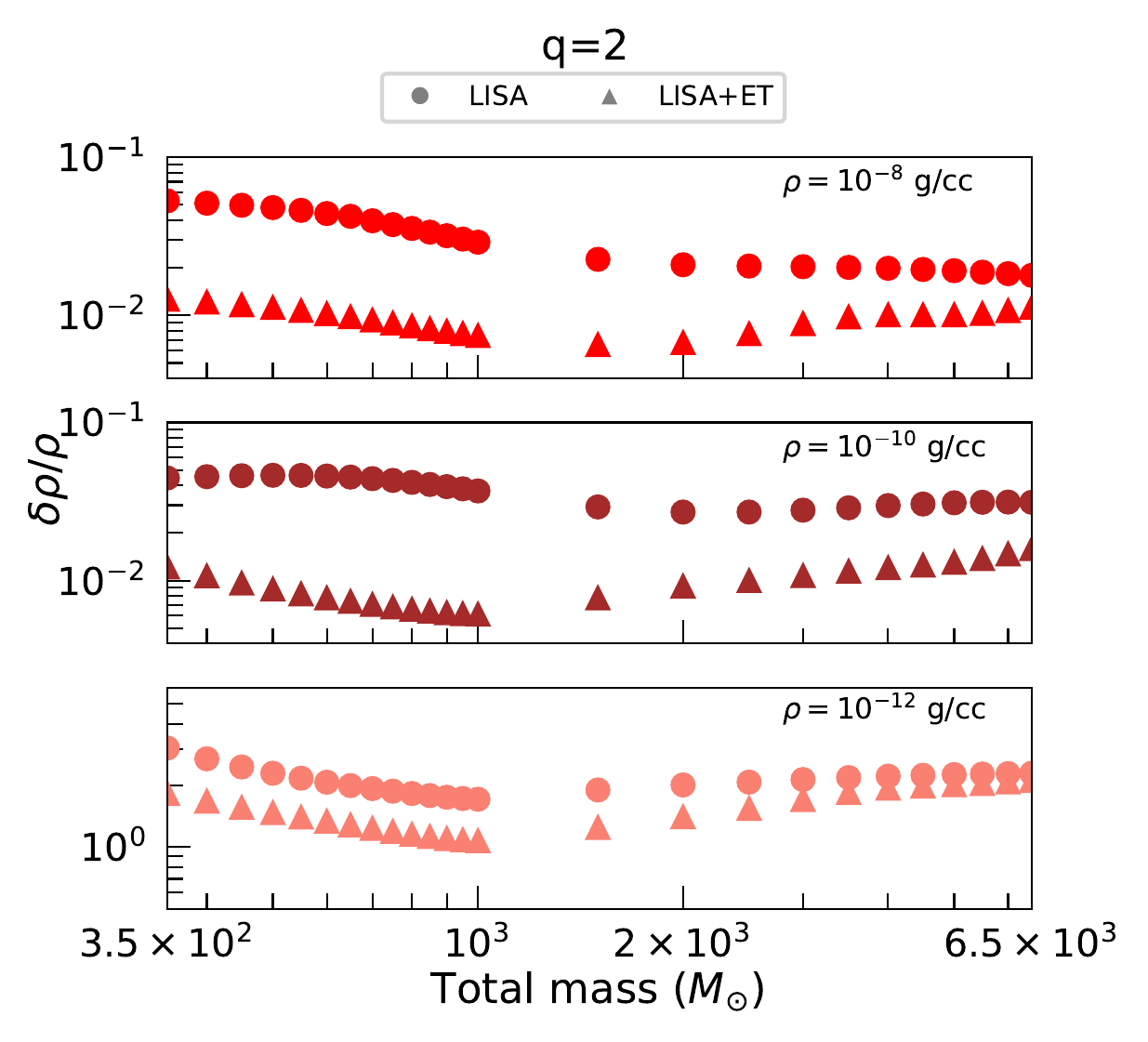}%
    }
    \hspace{0.3in}
   \subfloat[]{\label{fig:bound_q}%
  \includegraphics[width=0.41\textwidth]{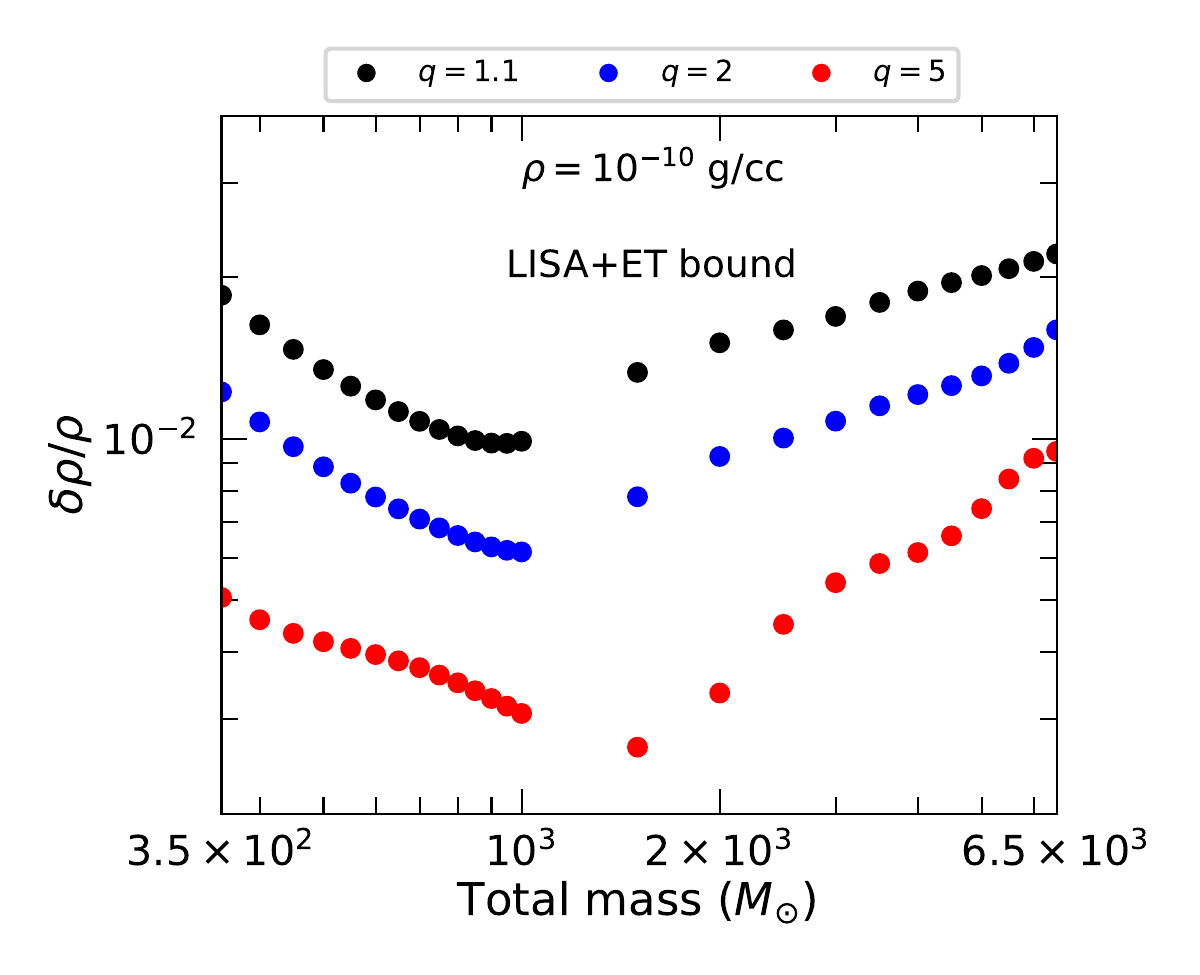}%
} 
    \caption{(a) The plot shows the fractional error on density for $\rho =10^{-8},\,10^{-10}$, and $10^{-12}$ g/cc obtained from LISA-only observations (circles) and multibanding with LISA and ET using {\tt IMRPhenomD} (triangles). The scaling factor is fixed at $\rho_0= 10^{-10}$ g/cc. All IMBBHs are observed in LISA with a 4-year observation time at a luminosity distance of 3 Gpc, with $q=2$ and aligned spins (0.2,0.1). The fractional error improves with increasing value of true density. (b) The plot shows the multibanding fractional error on density for mass ratios $q=2,5$, assuming a true density of $\rho = 10^{-10}$ g/cc and the scaling factor $\rho_0= 10^{-10}$ g/cc. The bound improves for more asymmetric systems.}
    \label{fig:bound_density}
\end{figure}
We now present our results for the measurement uncertainties associated with the density parameter from IMBBH mergers in LISA, as well as systematic biases that arise from neglecting the environmental effects in the inference of different parameters.

\subsection{Projected error bars on the density of the medium using LISA }\label{sec:LISA_error}
We begin by examining the precision with which the environmental effects can be measured using IMBBH mergers in LISA, followed by the effect of multibanding on these estimates. We assume three representative values for the true density of the environment, $10^{-8}, \,10^{-10}$, and $10^{-12}$ g/cc, which represent the typical average densities of AGN disks~\cite{Fabj:2024kqs,Vajpeyi:2021qsw}.

In Fig.(\ref{fig:bound}), we consider IMBBHs with a total mass ranging between $350-6500 M_\odot$ with a mass ratio $q=2$ and spins aligned with respect to the orbital angular momentum vector, with magnitudes $\chi_1=0.2, \chi_2=0.1$. We assume the sources to be at a fiducial luminosity distance of 3 Gpc. The chosen mass range ensures that the SNR of binaries in both LISA and ET, which  serves as our representative for the next-generation ground-based observatory, is centered around $\sim\mathcal{O}(100)$. Note that the dimensionless density parameter $\bar\rho$ was introduced in order to have a well-behaved Fisher matrix. To visualise the results in a more physically meaningful way, we translate the $1-\sigma$ error bars on the scaled density $(\Bar{\rho})$  to the fractional error on  the density which is shown in the top plot of Fig.(\ref{fig:bound}).The $\rho$ in the denominator of $\delta\rho/\rho$ corresponds to the particular density being considered.

For medium with $\rho=10^{-8} \, {\rm g/cc}$ and $10^{-10}\, {\rm g/cc}$, the relative errors are of the order of a few percent, implying that LISA will be able to place very tight constraints on the properties of the medium in these cases. Even for densities of $\sim$ $10^{-12}{\rm g/cc}$, a relative error ${\cal O}(100\%)$ would be very useful in understanding the non-vaccuum environment of the merger and would suffice to identify the merger sites to be AGN disks, which are the most natural candidates to host environments with densities of this order. Multibanding improves these errors by a factor of $2-5$ for those IMBBHs that can be observed in both bands. In the mass range we consider, the best measurement achieves an error bar of $\sim 0.6\%$ for  a $1000M_{\odot}$ binary with $q=2$ with multibanding. 
Figure (\ref{fig:bound_q}) shows the dependence of the errors on the mass ratio for $\rho =10^{-10}$ g/cc. As the mass ratio increases, the fractional errors decrease. This behavior can be attributed to the strong dependence of the dynamical friction term in the phase on the symmetric mass ratio (see Eq.~(\ref{eq:DF})). 
The non-monotonic nature of the curves in Fig.~\ref{fig:bound_q} can be attributed to multibanding of the low- and high- frequency data of LISA and ET, respectively, which gives rise to a minimum around $10^3M_{\odot}$.
\begin{figure}[t]
   \centering
    \includegraphics[scale=0.4]{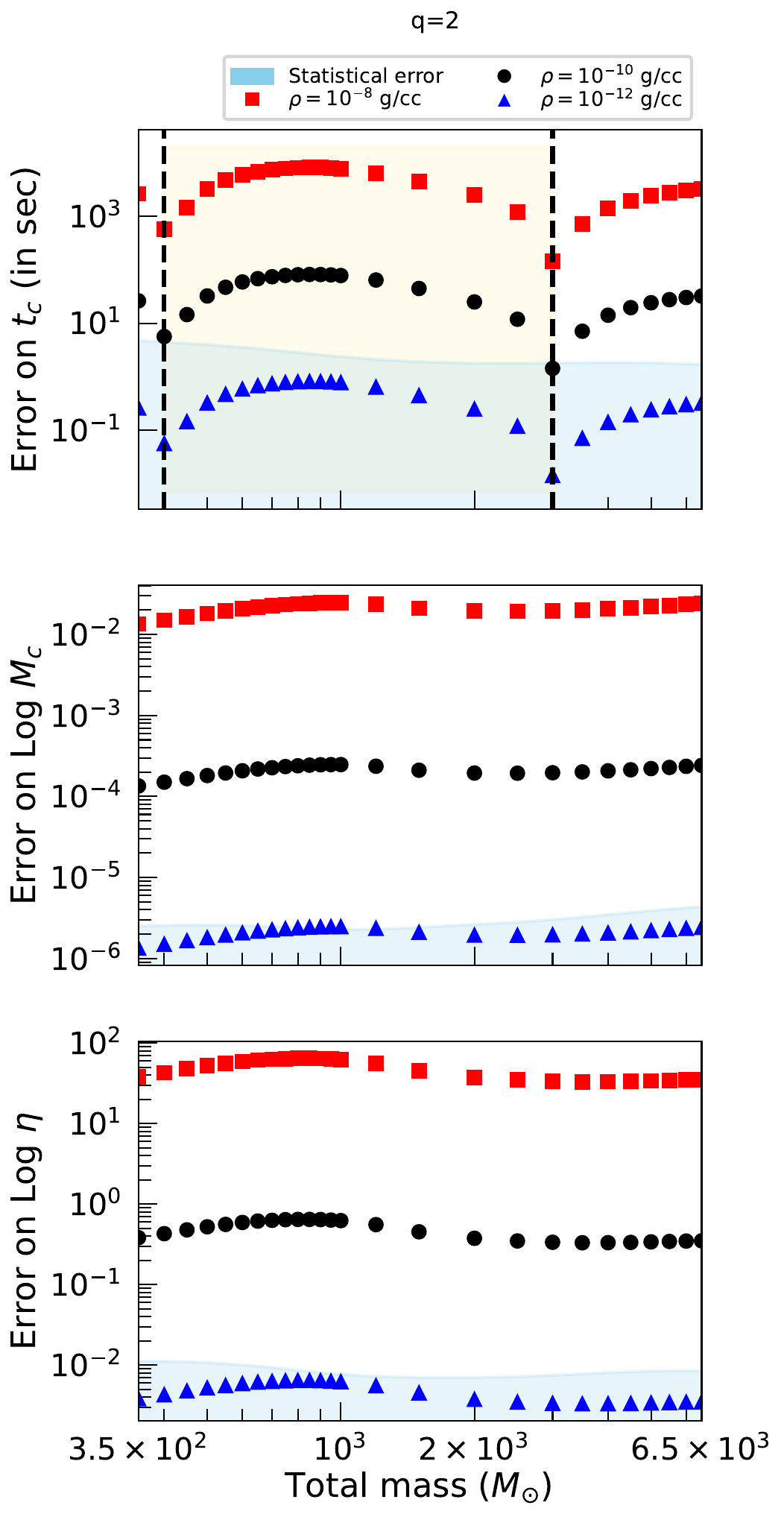}
    \caption{The plot shows the variation of absolute value of systematic error on $\rm t_c$, $\log \rm M_c$, and $\log \,\rm \eta$ due to dynamical friction occurring at $-$5.5PN, for different values of density and total mass. The corresponding statistical error is shown by the boundary of the blue-shaded region. The yellow-shaded region in the plot for $\rm t_c$ denotes the systematic error being positive while for the rest of the parameter space, the systematic error is negative. All IMBBHs are observed in LISA with a 4 year observation time at a luminosity distance of 3Gpc having $q=2$ and aligned spins (0.2,0.1) with $\rho_0=10^{-10}$ g/cc. 
    }
    \label{fig:syserr}
\end{figure}

\subsection{Systematic biases due to the neglect of environmental effects}

We now investigate the implications of neglecting environmental effects, such as dynamical friction, while performing parameter estimation with LISA. Since the phase contribution due to dynamical friction depends on the chirp mass and the symmetric mass ratio, ignoring this $-$5.5PN term is expected to impact the estimation of these parameters. Moreover, for LISA, IMBBHs undergo a large number of inspiral cycles in the detector sensitivity band, leading to the accumulation of systematic biases.

Here, we focus on the biases in the estimation of the time of coalescence $\rm{t_c}$, chirp mass $\rm \log\, M_c$, and symmetric mass ratio $\log \eta$. As mentioned earlier, Cutler--Vallisneri formalism is employed to estimate the systematic errors. Fig.(\ref{fig:syserr}) illustrates the variation in systematic error in $\rm{t_c}$, $\rm \log\, M_c$, and $\rm \log\,\eta$ for IMBBHs with mass ratio $q=2$ at a luminosity distance of 3 Gpc,  with aligned spin magnitudes $\chi_1=0.2, \chi_2=0.1$, for the three different densities considered earlier. The blue-shaded region in all three plots represents the statistical errors.
As expected, the systematic bias increases with the true average density.

Focusing first on $\rm{t_c}$, we notice a change in the sign of the systematic error, denoted by the two dashed vertical lines in Fig.~\ref{fig:syserr}. The systematic error is positive for masses within the yellow-shaded region, while for the rest of the explored  mass range, it is negative. This variation in the nature of the systematic bias on $\rm{t_c}$ is attributed to the correlations between density and $\rm{t_c}$, which change sign as a function of total mass. 
The statistical uncertainty associated with the measurement of $\rm{t_c}$ is of the order of $\mathcal{O}(10^{-1} - 1\,s)$ for binaries in AGN disks with density of $\sim 10^{-12}{\rm g/cc}$. However, it can be as large as $\mathcal{O}(10^2 - 10^3\,s)$  for densities of $\sim 10^{-10}{\rm g/cc}$ and $\sim 10^{-8}{\rm g/cc}$.

On the other hand, for the chirp mass, the statistical uncertainty in the $\rm \log\, M_c$ measurement is $\sim 10^{-6}$ for all binaries considered but the systematic bias, even for an environmental density of $10^{-12} \rm g/cc$, is of $\mathcal{O}(10^{-5})$ which should be a serious concern. It is observed that the systematic bias can be as high as $10^{-1}$ for higher densities. A similar trend is  also observed for $\log \eta$, where for densities above $10^{-10} {\rm g/cc}$, the systematic biases can be two orders of magnitude larger than the statistical errors. This highlights the critical issue of systematic bias due to the neglect of environmental effects in measuring the chirp mass when an IMBBH is detected in LISA.

Besides biasing the parameter inference, neglecting  environmental effects can significantly impact the efficiency of multiband GW observations. When the early inspiral phase of an IMBBH is first observed in LISA, the merger time and the corresponding arrival time of the GW signal in the frequency band of the ground-based detectors can be estimated. However, since the parameters $\rm{t_c}$ and $\rm M_c$ are biased, the corresponding estimation of the merger time and arrival time will also be significantly biased. This could lead to an erroneous estimation of signal arrival time from a particular binary in the ground-based detector band,  such as ET, thereby affecting the detection and parameter estimation of such multiband events. On the other hand, since ground-based detectors are insensitive to low-frequency environmental effects, the parameter estimation of a particular binary as performed using signals from ground-based detectors will be relatively unbiased. However, using these parameters  to perform an archival search in LISA observation data might also be problematic, since the parameters estimated from LISA and ET for the same binary may not match with each other. 
Furthermore, these biases will naturally propagate into any fundamental physics that rely on combining LISA and ground-based detectors~\cite{Gupta:2024gun}. Given these findings, it is evident that accounting for environmental effects in the LISA parameter estimation is crucial for several reasons.

\section{Conclusions}

X-ray observations have traditionally been used to infer the densities of AGN disks. These measurements are challenging both from an observational point of view and from a modeling point of view. Our results show that gravitational wave observations provide a new method, complementary to traditional approaches, for measuring the densities of AGN disks with exquisite precision, provided the local densities are higher than $10^{-12} \, {\rm g/cc}$. LISA observations of IMBBHs in AGN disks would carry signatures of the environment (in this case, a gas-rich medium) due to the modification to the low-frequency part of the phasing caused by dynamical friction. This effective $-$5.5PN term will enable us to confidently establish the merger site via measurement of the gas density with percent-level precision if the densities are greater than $10^{-12} \, {\rm g/cc}$. The measurement precision will improve further if a subset of the LISA IMBBHs is also observed in the frequency band of next-generation terrestrial GW detectors. Finally, we have shown that neglect of the environmental effects can significantly bias the estimation of chirp mass, mass ratio, and time of coalescence, which in turn will impact the prospects of carrying out multi-band observations of these sources. Therefore, it is critical that the possibility of mergers observed by LISA, occurring in nonvacuum environment, be incorporated into LISA's data analysis.

\section{Acknowledgments}
We acknowledge useful discussions with A. Gupta and B. Sathyaprakash. P.M. acknowledges the Science and Technology Facilities Council (STFC) for support through grant ST/V005618/1.
 K.G.A.~acknowledges the support of the Core Research Grant CRG/2021/004565 and the Swarnajayanti Fellowship Grant DST/SJF/PSA-01/2017-18 of the Science and Engineering Research Board of India. P.D.R and K.G.A acknowledge a grant from the Infosys Foundation.
\bibliographystyle{apsrev}
\bibliography{ref}
\end{document}